\documentclass[useAMS]{mn2e}
\usepackage{color}
\usepackage{graphicx}
\usepackage{amsmath}

\title[Cep X-4 with Suzaku]
{Detection of fundamental and first harmonic cyclotron line in X-ray pulsar Cep~X-4}
\author[G. K. Jaisawal and S. Naik]{Gaurava K. Jaisawal\thanks{gaurava@prl.res.in} and Sachindra Naik\thanks{snaik@prl.res.in} \\
Astronomy and Astrophysics Division, Physical Research Laboratory, Navrangapura, Ahmedabad - 380009, Gujarat, India\\}

\begin{document}

\date{}

\maketitle

\begin{abstract}

We report the broad-band spectral properties of the X-ray pulsar 
Cep~X-4 by using a {\it Suzaku} observation in 2014 July. The 0.8-70 keV
spectrum was found to be well described by three continuum models - Negative 
and Positive power-law with Exponential cutoff (NPEX), high energy cutoff power-law 
and CompTT models. Additional components such as a cyclotron line at $\sim$28~keV 
and two Gaussian components for iron lines at 6.4 and 6.9~keV were required in the 
spectral fitting. Apart from these, an additional absorption feature at $\sim$45~keV 
was clearly detected in residuals obtained from the spectral fitting. 
This additional feature at $\sim$45~keV was clearly seen in phase-resolved 
spectra of the pulsar. We identified this feature as the first harmonic of 
the fundamental cyclotron line at $\sim$28~keV. The ratio between the first 
harmonic and fundamental line energies (1.7) was found to be in disagreement 
with the conventional factor of 2, indicating that the heights of line forming 
regions are different or viewed at larger angles. 
The phase-resolved spectroscopy of the fundamental and 
first harmonic cyclotron lines shows significant pulse-phase variation of the 
line parameters. This can be interpreted as the effect of viewing angle or the 
role of complicated magnetic field of the pulsar.     
     
\end{abstract}

	 
\section{Introduction}

Cyclotron resonance scattering features (CRSFs) are generally seen in the 
hard X-ray spectrum (10-100~keV) of the accretion powered X-ray pulsars with
surface magnetic field of $\sim$10$^{12}$ G. These are absorption like features 
which appear due to the resonant scattering of photons with electrons in quantized 
Landau levels (M{\'e}sz{\'a}ros 1992). The energy difference between these levels 
depends on the strength of magnetic field and expressed through the relation 
{\it E$_{cyc}$=11.6B$_{12}\times(1+z){^{-1}}$ (keV)} (without relativistic 
correction); where B$_{12}$ is the magnetic field in the unit of 10$^{12}$~G 
and $z$ is the gravitational red-shift. Detection of fundamental CRSF in the 
spectra of accretion powered X-ray pulsars provides the direct estimation of 
local magnetic field of the neutron stars in line forming region. 
At the same time, the study of the harmonics (multiples of fundamental 
cyclotron line) gives crucial information about the optical depth of the 
line-forming region (Araya-G{\'o}chez \& Harding 2000; Sch{\"o}nherr et al. 2007; 
Nishimura 2013). As of now, CRSF has been seen in about 20 accretion powered X-ray 
pulsars. However, the harmonics of the fundamental cyclotron line are detected only 
in a few cases (Pottschmidt et al. 2012 and references therein).

The transient X-ray pulsar Cep~X-4 (GS~2138+56) was discovered at a position 
near the galactic plane with {\it OSO-7} in 1972 (Ulmer et al. 1973). X-ray 
pulsations were detected in the source at $\sim$66~s during its 1988 March 
outburst with {\it Ginga} (Makino et al. 1988; Koyama et al. 1991). A cyclotron
absorption line feature at $\sim$30~keV was discovered in 1.2-37~keV range spectrum 
obtained from $Ginga$ observation of the pulsar (Mihara et al. 1991). An optical 
star of 14.2 magnitude was identified as the companion of Cep~X-4 (Roche, Green 
\& Hoenig 1997). Detailed optical spectroscopy showed the presence of Balmer 
emission lines that characterized the companion as Be star of type B1-B2V. The 
distance of the binary was estimated as 3.8~kpc (Bonnet-Bidaud \& Mouchet 1998). 

The presence of cyclotron line at $\sim$30~keV was confirmed with the
2002 June $RXTE$ observations of Cep~X-4 (McBride et al. 2007). However, 
there was no significant changes in cyclotron line energy with source 
luminosity observed during these observations. The pulsar was observed
with {\it NuSTAR} during the 2014 June-July outburst. First observation
was carried out near the peak of the outburst whereas the second was at
the declining phase of the outburst (F{\"u}rst et al. 2015). The 1-50~keV 
combined spectra from {\it Swift/XRT} and {\it NuSTAR} were described 
by Fermi-Dirac cutoff power-law (FDCUT) model along with an absorption 
component at $\sim$30~keV. Addition of a simple absorption component or 
a pseudo-Lorentzian profile was failed to explain the cyclotron absorption 
feature properly because of the asymmetric nature of the line profile. 
Along with this asymmetry profile, an absorption like feature at $\sim$19~keV 
was also detected in the spectrum. A marginal positive dependence of cyclotron 
line energy with the pulsar luminosity was seen during the {\it NuSTAR} 
observations.    
    
In the present work, we used {\it Suzaku} observation of the pulsar 
during 2014 outburst to study its spectral properties. The broad-band 
coverage and low background capability (up to 70~keV) of detectors onboard 
{\it Suzaku} provided best opportunity to investigate cyclotron absorption
line parameters in binary X-ray pulsars. Using {\it Suzaku} observation
of Cep~X-4, we detected a cyclotron line at $\sim$28~keV along with another 
absorption like feature at $\sim$45~keV. We interpret the absorption lines 
at $\sim$28~keV and at $\sim$45~keV as the fundamental and first harmonic 
cyclotron absorption lines, respectively. Details on the observations, 
analysis procedures, results and conclusions are presented in following 
sections of the paper.

\section{Observation and Analysis}

A Target of Opportunity (ToO) observation of Cep~X-4 was carried out 
with {\it Suzaku} (Mitsuda et al. 2007) in July 01-02 during its 2014 June-July 
outburst. The pulsar was observed for exposures of $\sim$60~ks with XIS-3 and 
$\sim$81~ks with HXD during the decay phase of the outburst. Among three 
active XISs (XIS-0, 1 \& 3), data from the XIS-3 was used in our analysis as
XIS-0 and XIS-1 were exposed for $\sim$100~s short durations. The XIS-3  
was operated in ``normal'' clock mode with ``1/4'' window option 
yielding 2~s time resolution. The observation was carried out in 
``XIS nominal'' position. We used publicly available data (Obs. ID: 
909001010) in the present study. Calibration database (CALDB) files 
released on 2015 January 05 (XIS) and 2011 September 13 (HXD) were 
applied during reprocessing of data in Heasoft (version 6.16) 
analysis package.

The `aepipeline' task of FTOOLS was used to reprocess the unfiltered 
XIS and HXD events. Clean events generated after the reprocessing 
were used in our study.  Barycentric correction was applied on these 
XIS and PIN clean events by using `aebarycen' package. XIS data were 
checked for the attitude and pile-up effects by applying S-lang scripts 
(\textit{aeattcor.sl} and  
\textit{pile\_estimate.sl}), 
respectively. We found a pile-up of $\sim$8\% at the centre of XIS-3 that 
was reduced to $\le$4\% by choosing the events only from an annulus region 
with inner and outer radii of $10\arcsec$ and $180\arcsec$, respectively. The source 
light-curves and spectra from XIS-3 were extracted by using {\it XSELECT} 
package of FTOOLS. Background light-curves and spectra for XIS-3 were created 
from a circular region away from the source. Response matrix and effective area 
files for XIS-3 were generated from ``resp=yes" command in {\it XSELECT}. Source 
light curves and spectra for PIN were extracted from cleaned event data by using 
{\it XSELECT}. However, HXD/PIN background light curves and spectra were accumulated 
in a similar manner from  simulated non-X-ray background event file provided by
the instrument team.  The cosmic X-ray background correction 
was included in PIN non-X-ray background spectrum. Epoch 11 PIN response 
file (20110601) was used in spectral analysis.

\section{Results}

\begin{figure}
\centering
\includegraphics[height=2.1in, width=1.8in, angle=-90]{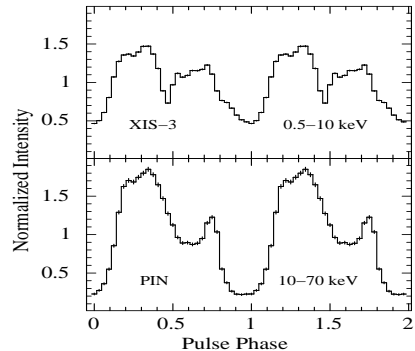}
\caption{Pulse profiles of Cep~X-4 obtained from XIS-3 and HXD/PIN light curves
during the $Suzaku$ observation in 2014 July. The presence of absorption dip in 
soft X-ray pulse profile in 0.4-0.5 phase range can be seen. The error bars represent 
1$\sigma$ uncertainties. Two pulses in each panel are shown for clarity.}
\label{pp}
\end{figure}

Source and background light curves with 2~s and 1~s time resolution were 
extracted from the barycentric corrected XIS-3 and PIN event data, respectively. 
The $\chi^2$-maximization technique was used to estimate the pulse period of the 
pulsar. The pulsation was detected at a period of 66.334$\pm$0.004~s from  
background subtracted light curves of XIS-3 and PIN. Above pulsation period was 
used to generate pulse profiles in soft (0.5-10~keV) and hard X-rays (10-70~keV)
and are shown in Fig.~\ref{pp}. Strong energy dependence of pulse profiles with 
energy can be seen in the soft and hard X-rays pulse profiles. A dip like feature 
in soft X-rays (top panel) in 0.4-0.5 phase range disappeared from the hard X-ray 
pulse profile. Therefore, it is interesting to investigate spectral properties of 
the pulsar at different pulse phases.

\subsection{Spectral Analysis}

\subsubsection{Pulse-phase-averaged spectroscopy}

To study the broad-band spectral characteristics of the pulsar, 
phase-averaged spectroscopy was carried out by using the source 
and background spectra accumulated from the XIS-3 and PIN event 
data. The procedure for spectral extraction was described earlier. 
The 0.8-70~keV spectra, obtained from XIS-3 and PIN data were 
simultaneously fitted by using $XSPEC$ (ver. 12.8.2) package. 
Appropriate background spectra, response matrices and effective area 
files for corresponding detectors were used in the spectral fitting. 
Spectral data in 1.7-1.9~keV and 2.2-2.4~keV ranges were ignored in 
the fitting due to presence of known artificial emission features in 
the spectrum. XIS-3 spectrum was binned by a factor of 6, whereas PIN 
spectrum was binned by a factor of 2 up to 30~keV, a factor of 4 from 
30 to 50~keV and a factor of 6 from 50 to 70~keV. All the spectral 
parameters were tied during the fitting, except the normalization 
constant of detectors which were kept free.

\begin{figure}
\centering
\includegraphics[height=2.9in, width=2.3in, angle=-90]{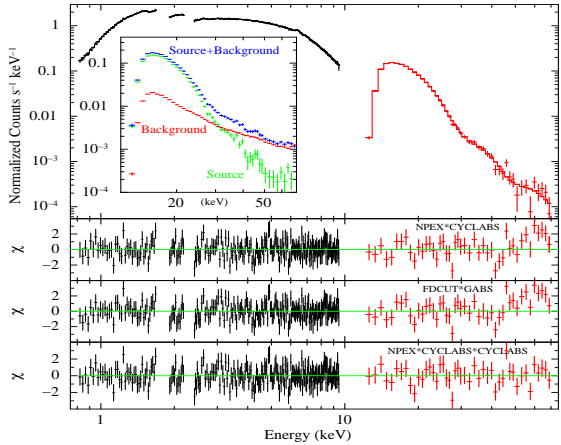}
\caption{Energy spectrum of Cep~X-4 in 0.8-70~keV range obtained from XIS-3 
and HXD/PIN data, along with the best-fit model comprising a partial covering 
NPEX model, two Gaussian functions for iron emission lines and two cyclotron 
absorption components.  The second and fourth panels show the contributions 
of the residuals to $\chi^{2}$ for each energy bin for the partial covering NPEX 
continuum model with one and two cyclotron absorption components, 
respectively. The third panel shows the residuals for FDCUT model with one GABS component.   
The inset shows the HXD/PIN spectrum of the pulsar with background,
without background and the simulated HXD/PIN background spectrum.}
\label{npex}
\end{figure}

\begin{table}
\centering
\caption{Best-fitting parameters (with 90\% errors) obtained from the spectral fitting 
of $Suzaku$ observation of Cep~X-4 in 2014 July. Model-1 \& Model-2  
consist of the partial covering NPEX and CompTT model with 
two Gaussian and two cyclotron lines, respectively. Model-3 is combination of FDCUT 
model with blackbody, two Gaussian and one $gabs$ components.}

\begin{tabular}{llll}
\hline
Parameter                       &Model-1             &Model-2            &Model-3 \\
\hline
N$_{H1}$$^a$                    &0.78$\pm$0.03       &0.77$\pm$0.03     &0.71$\pm$0.02 \\
N$_{H2}$$^b$                    &3.7$\pm$0.6         &2.6$\pm$0.3       &--       \\
Cov. fraction                   &0.28$\pm$0.06       &0.41$\pm$0.08     &--     \\
Photon index                    &1.04$\pm$0.09       &--                &1.01$\pm$0.05 \\
BB temp. (keV)                  &--                  &--                &1.08$\pm$0.06  \\
BB norm. (10$^{-4}$)              &--                  &--                &4.8$\pm$0.6 \\
E$_{cut}$ (keV)	                &9.5$\pm$2.1         &--                &25.1$\pm$4.0   \\
E$_{fold}$ (keV)                &--                  &--                &6.8$\pm$1.1    \\
CompTT$_{T0}$ (keV)             &--                  &0.21$^{+0.6}_{-0.21}$  &--         \\    
CompTT ${kT}$ (keV)             &--                  &11.6$\pm$2.3      &--              \\
CompTT ${\tau}$                 &--                  &4.4$\pm$0.4       &--              \\

{\it Fe line parameters} \\
Energy (keV)                    &6.41$\pm$0.03       &6.41$\pm$0.03    &6.42$\pm$0.03 \\
Eq. width (eV)                  &32$\pm$8            &27$\pm$7         &34$\pm$8  \\

Energy (keV)                    &6.97$\pm$0.16       &6.98$\pm$0.22    &6.95$\pm$0.14 \\
Eq. width (eV)                  &13$\pm$6            &9$\pm$6          &14$\pm$7\\

{\it Cyc. line parameters} \\
Energy (E$_{c1}$) (keV)         &27.5$\pm$0.4      &27.7$\pm$0.4     &29.6$\pm$0.5$^*$    \\
Width ($\sigma_{c1}$) (keV)     &8.9$\pm$1.0       &6.1$\pm$0.8      &4.8$\pm$0.5$^*$      \\
Depth ($D_{c1}$)	              &2.3$\pm$0.3       &1.8$\pm$0.3      &17$^{+5}_{-3}$$^*$      \\
Energy (E$_{c2}$) (keV)         &45.4$\pm$2.8      &43.0$\pm$3.6     &--       \\
Width ($\sigma_{c2}$) (keV)     &10.3$\pm$4.6      &11.9$\pm$4.2     &--         \\
Depth ($D_{c2}$)	              &1.8$\pm$0.7       &1.8$\pm$0.4      &--       \\

Flux$^c$ (1-10 keV)             &3.1$\pm$0.5       &3.1$\pm$0.7    &3.1$\pm$0.4 \\
Flux$^c$ (10-70 keV)            &5.5$\pm$1.1       &5.6$\pm$1.2    &5.5$\pm$1.4\\ 
$\chi^2$ (dofs)                 &254 (244)         &261 (244)      &303 (247) \\
\hline
\end{tabular}
\\
\flushleft
$^a$ : Equivalent hydrogen column density and $^b$ : Additional hydrogen column density (in 10$^{22}$ atoms cm$^{-2}$); 
$^c$ : Absorption uncorrected flux (in 10$^{-10}$  ergs cm$^{-2}$ s$^{-1}$.); $^*$ : cyclotron parameters are 
given for $gabs$ component. \\

\label{spec_par}
\end{table}

Continuum models used to describe spectra of accretion powered X-ray pulsars 
such as high energy cutoff power-law (White et al. 1983), cutoff power-law, 
NPEX (Makishima et al. 1999), FDCUT and CompTT (Titarchuk 1994) were used to 
fit the 0.8-70 keV spectrum of Cep~X-4. Investigating the residuals, we added 
two Gaussian functions at 6.4 and 6.9~keV for iron emission lines in the source
spectrum. As in case of Be/X-ray binary pulsars, a partial covering absorption
component (Paul \& Naik 2011 and references therein) was used in the spectral
model along with the interstellar absorption component. A strong absorption like 
feature at $\sim$28~keV was clearly seen in the spectrum. Addition of a CRSF
component at above energy was added to all continuum models. Simultaneous 
spectral fitting of XIS-3 and HXD/PIN data in 0.8-70 keV range showed that  
partial covering NPEX continuum model, partial covering high energy cutoff 
power-law model and partial covering CompTT model describe the spectrum well 
with acceptable values of reduced $\chi^2$ ($<$1.5). 

Apart from $\sim$28~keV cyclotron line, we found an additional absorption like 
feature at $\sim$45~keV in the residues obtained from all three continuum models. 
This feature was clearly seen at same energy range and was model independent 
(third panel in Fig.~\ref{npex}). The inset in Fig.~\ref{npex} shows the 
HXD/PIN spectrum of the {\it Suzaku} observation of the pulsar with background, 
without background and simulated PIN background spectrum. The absorption-like 
feature at $\sim$45 keV can be clearly seen in the background subtracted 
spectrum. The inclusion of additional CRSF component at this energy in the
model fitted the XIS-3 and HXD/PIN spectra well with significant improvement 
in the values of $\chi^2$. Absorption feature at $\sim$45~keV can be  
considered as the first harmonic of the $\sim$28~keV cyclotron absorption line. 
Ideally, the ratio between the energy of first cyclotron harmonic and fundamental 
line is expected to be a value closer to 2. However, in present case, the ratio 
was estimated to be $\sim$1.7$\pm$0.1, which can be acceptable in current 
understanding of the cyclotron physics. Best-fit parameters obtained from 
simultaneous spectral fitting are given in Table~1. The energy spectra of 
the pulsar along with all three best-fitted model components resemble similar 
absorption features in the spectral residues. Energy spectra obtained from 
simultaneous fitting of the XIS-3 and HXD/PIN data are shown in Fig.~\ref{npex}. 
The second and fourth panels in figure show the residuals to the best-fit 
model with one and two cyclotron lines in the continuum model, 
respectively. As in case of {\it NuSTAR} observations, a weak absorption like 
feature can be seen at $\sim$19~keV in the spectral residue (third and fourth 
panels of Fig.~\ref{npex}). Addition of an absorption component at $\sim$19~keV, 
however, did not show any significant improvement in the spectral fitting. 
Negligible strength and width of the $\sim$19~keV line against other two 
absorption lines at $\sim$28 and $\sim$45~keV, therefore, makes it difficult 
to accept the earlier feature as the fundamental cyclotron absorption line. 
Therefore, we did not include any spectral component for this feature in our 
fitting. 

To verify the $\sim$45 keV feature as additional absorption feature detected 
in {\it Suzaku} spectra, we fitted XIS-3 and HXD/PIN spectra with the spectral model 
(FDCUT model along with black-body and Gaussian absorption (gabs) components) as 
used by F{\"u}rst et al. (2015) to fit spectra obtained from {\it Swift} and {\it NuSTAR} 
observations of the source. Though the data shown in Fig.~2 of F{\"u}rst et al. (2015) 
were up to 50 keV, we used {\it Suzaku} data in 0.8-70~keV range to compare both the 
results. The spectral parameters obtained from our fitting were found to be comparable 
(within errors) with the parameters obtained from the {\it NuSTAR} observations  
and given in Table~1. While fitting {\it Suzaku} data with the model used by 
F{\"u}rst et al. (2015), an absorption like feature at $\sim$45~keV was also seen 
(third panel of Fig.~\ref{npex}). A careful look at the trend of distribution 
of points at $\sim$45-50 keV in the residuals obtained from the spectral fitting of 
Obs.1 data (second and third panels of Fig.~2 of F{\"u}rst et al. (2015)) indicates 
a hint of presence of an absorption-like feature. 

Statistical significance of $\sim$45~keV absorption feature was tested by 
using {\it XSPEC} script $simftest$ as applied in case of IGR~J17544-2619 (Bhalerao 
et al. 2015). Including systematic uncertainty of 0.3\% (15-40~keV) and 1.9\% (40-70~keV) 
in PIN background model, we simulated 1000 fake spectra for partial covering NPEX model 
and estimated the differences in $\chi^2$ ($\varDelta\chi^2$) without and with $\sim$45~keV 
cyclotron absorption line. The maximum value of $\varDelta\chi^2$ from 1000 simulations 
was found to be 17.8 which is less than observed $\varDelta\chi^2$=27.2 for three degree of 
freedom in real data. Corresponding to this, we confirmed the significant detection of 
$\sim$45~keV absorption feature at $>4\sigma$ level.

\begin{figure}
\centering
\includegraphics[height=2.8in, width=1.9in, angle=-90]{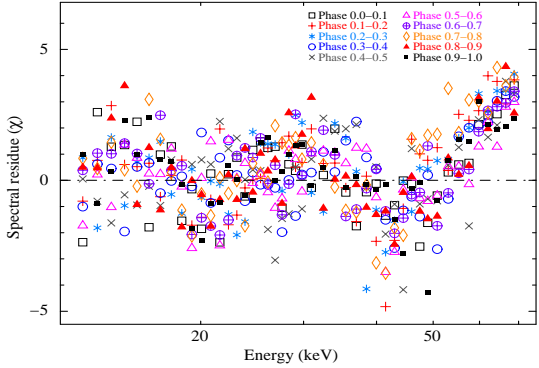}
\caption{Spectral residuals obtained from the phase-resolved spectra while fitted with 
partial covering NPEX continuum model with a fundamental cyclotron absorption line at
$\sim$28 keV. An absorption like feature in $\sim$40-50~keV range is clearly seen in 
each of the phase intervals of the pulsar.}
\label{spec_res}
\end{figure}

\begin{figure}
\centering
\includegraphics[height=2.2in, width=2.9in, angle=-90]{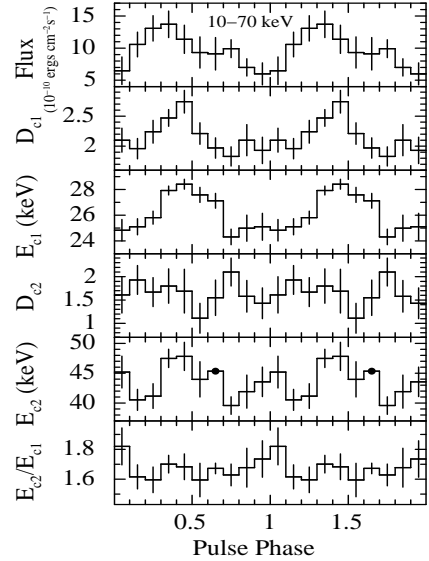}
\caption{Spectral parameters (with 90\% errors) obtained from the phase-resolved 
spectroscopy of Cep~X-4. Top panel shows estimated source flux in 10-70~keV energy
range. The values of cyclotron line parameters such as depths and energies for fundamental 
and first harmonic and ratio between harmonic and fundamental line energies (E$_{c2}$/E$_{c1}$) 
are shown in second, third, fourth, fifth and sixth panels, respectively. Solid circles in 
the fifth panel indicate that the cyclotron line energy was fixed for corresponding phase-bin 
at the phase-averaged value.}
\label{spec_para}
 \end{figure}

\subsubsection{Pulse-phase-resolved spectroscopy}

We performed phase-resolved spectroscopy to understand the change in cyclotron 
absorption line parameters with the pulse phase of the pulsar. Another motivation 
in doing phase-resolved spectroscopy was to study the presence of $\sim$45~keV 
absorption line as well as its dependence on pulse-phases of the pulsar. Detection 
of absorption feature at $\sim$45 keV at each pulse-phase bins can confirm the 
detection of first cyclotron harmonic in Cep~X-4. For this, the phase-resolved 
spectroscopy was carried out by accumulating source spectra in 10 pulse-phase bins. 
XIS-3 and PIN phase-resolved spectra were extracted by applying phase filter in 
{\it XSELECT}. Using the same background, response and effective area files as 
used in phase-averaged spectroscopy, the phase-resolved spectroscopy was carried 
out in 0.8-70~keV range. Partial covering NPEX model with a cyclotron component at 
$\sim$28~keV was used to describe the phase-resolved spectrum. While fitting, the 
normalization constants, equivalent hydrogen column density (N$_{H_1}$), iron line 
parameters and cyclotron width were fixed to the corresponding phase-averaged values
as given in Table~1.

Spectral residues obtained from fitting all phase-resolved spectra with above model 
are shown in Fig.~\ref{spec_res}. Presence of an additional significant absorption 
feature in $\sim$40-50~keV range can be clearly seen in each phase intervals. Detection 
of the $\sim$45~keV absorption line in phase-averaged as well as phase-resolved spectra 
ensures that this feature is not spurious and model dependent. This feature can be 
interpreted as the first cyclotron harmonic of the $\sim$28~keV fundamental line. 
Inclusion of a CRSF at this energy improved the spectral fitting significantly
as in case of phase-averaged spectroscopy. During the fitting, widths of both 
cyclotron lines were frozen at phase-averaged value to constrain the absorption 
features. Variation of best fitted cyclotron line parameters with pulse phase 
are shown in Fig.~\ref{spec_para} along with hard X-ray source flux.

Cyclotron parameters such as depth and energy of both the lines were found 
significantly variable with pulse-phase with maximum values in 0.3-0.6 phase 
range. However, the peak values of these parameters were slightly phase shifted 
(0.1 phase) with the peaks of 10-70~keV pulse profile and source flux. Depth of 
fundamental line was variable in the range of 1.8 to 2.7 ($\sim$20\% of
the phase-averaged value). Depth of first cyclotron harmonic was found to be 
marginally variable with the pulse phase of the pulsar. The variation of the energy
of the first harmonic was found to be double ($\sim$8~keV) of that of the fundamental 
line ($\sim$4~keV). However, while computing the ratio between the energy of the 
first harmonic and the fundamental cyclotron line, it was found to be in $\sim$1.6-1.8 
range. Source flux in 10-70~keV range was found to follow similar pattern as 
the pulse profile in same energy band.

\section{Discussion and Conclusions}

We report the detection of first harmonic of the cyclotron absorption line in
Cep~X-4. Though the pulsar was observed with {\it NuSTAR} during same outburst, 
data from {\it Suzaku} observation with longer exposure (twice that of {\it NuSTAR}) 
confirmed the detection of the harmonic at $\sim$45 keV. The harmonic feature was 
found to be model independent and also present in each bin of the phase-resolved 
spectra of the pulsar. Statistical tests on the {\it Suzaku} data also confirmed
the detection of the harmonic of the cyclotron line at $\sim$45 keV. Though {\it 
NuSTAR} has better effective area at $\sim$50 keV than {\it Suzaku}, long exposure 
of the {\it Suzaku} observation detected the additional feature at $\sim$45 keV.

Cyclotron absorption features in the broad-band X-ray spectrum originate due to 
the resonance scattering of photons with quantized electrons in the presence of 
magnetic field. Depending on the strength of the magnetic 
field, the states of electrons are quantized in harmonically spaced levels such 
that the first harmonic energy is expected to be at twice of the fundamental 
energy. In the present case of Cep~X-4, however, the first harmonic is 
detected at an energy which is $\sim$1.7 times that of the fundamental 
line which is less than the ideal coupling factor of 2. An-harmonic spacing 
between fundamental and harmonic lines has also been seen in a few other X-ray 
pulsars and described by considering the relativistic effects in photon-electron 
scattering for small changes in the energy ratio (M{\'e}sz{\'a}ros 1992; 
Sch{\"o}nherr et al. 2007). However, this may not be the only 
cause that can produce the anharmonicity in lines. Cyclotron absorption phenomena 
for fundamental and harmonic lines occurring at two different scale heights can 
have different optical depths and introduce the anharmonicity in the coupling 
factor or line energy ratio. In detailed studies of cyclotron lines, Nishimura (2005) 
and Sch{\"o}nherr et al. (2007) showed that the increase in magnetic field within a line 
forming region can result the line ratio less than 2, as seen in 4U~0115+63 (Heindl et 
al. 1999) and Cep~X-4 (present work). At larger viewing angle $\mu$=0.79, the line 
ratio is expected to be 1.73 for polar cap radius of 1.5 km (Nishimura 2013). Such 
decrease in energy ratio (1.57-1.73) is possible for viewing angle of 0.52-0.79 
where superimposition of large numbers of fundamental line emerging from different 
heights of line-forming region shifts the energy to higher side with nearly constant 
energy for first harmonic. In another scenario, the anharmonicity in the line ratio 
can be expected due to distortion or displacement from the dipole geometry of the 
magnetic field. In this case, both CRSFs are generated at two different poles of 
neutron star and produce a significant phase shift between both line parameters. Such 
phase-shift was not seen in Cep~X-4 (Fig.~\ref{spec_para}).   

For the first time, we present the detailed analysis of fundamental cyclotron line 
with previously unknown first cyclotron harmonic in Cep~X-4. Numerous simulation works 
on cyclotron lines were done by considering certain sets of assumptions and geometry in 
line forming regions (Araya-G{\'o}chez \& Harding 2000; Sch{\"o}nherr et al. 2007; 
Mukherjee \& Bhattacharya 2012). These studies predicted that the cyclotron absorption 
line parameters are expected to show 10-20\% variation over pulse phases depending on 
the viewing angle of the accretion column. However, more than 30\% variation in cyclotron 
parameters can be explained by considering distortion in the magnetic dipole geometry of 
the pulsar (Sch{\"o}nherr et al. 2007; Mukherjee \& Bhattacharya 2012). In Cep~X-4, 
both cyclotron line parameters are varying within 20\% over pulse-phase, which can
be described in terms of the viewing angle or local distortion in magnetic field. 
Detailed modeling of the observed variations in cyclotron parameters would provide 
useful information about the neutron star magnetic field geometry, inclination 
and beaming or emission patterns. However, these works are beyond the scope of the 
paper. 

In summary, we report the detection of the first harmonic of $\sim$28 keV fundamental 
cyclotron absorption feature at $\sim$45~keV in Cep~X-4. This feature was clearly seen 
in the phase-averaged and phase-resolved spectra from the {\it Suzaku} observation in 
2014 July. The values of the energy of first harmonic with fundamental line were found 
anharmonic with ratio of 1.7. Parameters of both the fundamental and first harmonic 
lines were variable within 20\% with pulse-phase which is explained as the effect of 
viewing angle or local perturbation in magnetic field of line forming region.   

We sincerely thank the anonymous referee for valuable comments which improved the paper
significantly. This research has made use of $Suzaku$ data obtained through HEASARC Online Service.

\end{document}